\documentclass[aps,prd,eqsecnum,showpacs,nofootinbib]{revtex4}

\usepackage[dvipdfmx]{graphicx}
\usepackage{amsmath,amsfonts,amssymb,color,latexsym,theorem,mathrsfs,comment,color}

\usepackage{tikz}
\usetikzlibrary{matrix}
\usetikzlibrary{positioning}
\usetikzlibrary{intersections,calc}
\usetikzlibrary{arrows.meta}

\newcommand{\ma}[1]{\mbox{$\mathcal{#1}$}}
\newcommand{\mas}[1]{\mbox{$\mathscr{#1}$}}

\newcommand{\D}{{\rm d}}

\newcommand{\ti}{\tilde}

\begin{document}

\title{
Wormhole $C$ metric
}

\author{Masato Nozawa${}^1$ and Takashi Torii${}^2$}
\email{
masato.nozawa@oit.ac.jp, takashi.torii@oit.ac.jp
}

\address{
${}^1$General Education, Faculty of Engineering, 
Osaka Institute of Technology, Osaka City, Osaka 535-8585, Japan, \\
${}^2$Department of System Design, Osaka Institute of Technology, Osaka City, Osaka 530-8568, Japan
}

\date{\today}

\begin{abstract}
The C-metric in vacuum general relativity describes a pair of accelerated black holes supported by conical singularity. In this paper, we present a new family of exact solutions  to the Einstein-phantom scalar system that describes accelerated wormholes in AdS. In the zero acceleration limit with a vanishing potential, the present solution recovers the asymptotically flat wormhole originally constructed by Ellis and Bronnikov. The scalar potential of the phantom field has an infinite number of critical points  and is expressed in terms of the superpotential, which is obtained by suitable analytic continuation of one parameter family of the ${\cal N}=2$ gauged supergravity. As one traverses two asymptotic regions connected by throat, the scalar field evolves from AdS, corresponding to the origin of the potential, towards the neighboring AdS local minimum of the  potential. We find that the flipping transformation, which interchanges the role of ``radial'' and ``angular'' coordinates at the expense of double Wick rotation, is an immediate cause for the existence of two branches of static AdS wormholes discovered previously. Contrary to the ordinary C-metric, the conical singularity along the symmetry axis can be completely resolved, when the (super)potential is periodic or zero. We explore the global causal structure in detail. 
\end{abstract}

\pacs{} 

\maketitle

\section{Introduction}

A family of  vacuum solutions toEinstein's equations \cite{LeviCivita,Newman1961,Ehlers:1962zz}, which describes
a pair of causally disconnected black holes with a uniform acceleration \cite{Kinnersley:1970zw}, is dubbed as a C-metric.  
The vacuum C-metric is classified into Petrov type D and Weyl class of solutions \cite{Bonnor}. 
The acceleration of each black hole is produced by a conical 
deficit angle along the axis of symmetry, corresponding to the cosmic string extending out to infinity, or a strut with a negative tension 
developing between two black holes. At the linear approximation of acceleration, 
the C-metric appears as a perturbation of the Schwarzschild black hole with a distributional stringy source \cite{Kodama:2008wf}. 
Extensive studies have been conducted thus far on the causal structures and physical properties of the vacuum C-metric 
 \cite{Hong:2003gx,Letelier:1998rx,Griffiths:2006tk,Lim:2014qra}. 
The C-metric in anti-de Sitter (AdS) spacetime has also been investigated  from diverse perspectives, including 
causal structures 
\cite{Podolsky:2002nk,Dias:2002mi,Krtous:2005ej}, thermodynamics \cite{Appels:2016uha, Appels:2017xoe,Astorino:2016ybm,Zhang:2018hms,Wang:2022hzh},
minimal surfaces \cite{Xu:2017nut}, and 
quasi-normal modes \cite{Nozawa:2008wf,Destounis:2020pjk,Destounis:2022rpk}.

In our recent paper \cite{Nozawa:2022upa}, we have constructed a new family of C-metrics in ${\cal N}=2$ gauged supergravity. 
Upon suitable truncation, the bosonic part of this theory is nothing but the Einstein-${\rm U}(1)^2$-dilaton gravity. 
In the case of zero acceleration limit, this C-metric reduces to the asymptotically AdS, magnetically charged black hole with
a nontrivial scalar field, which admits a parameter range under which the event horizon persists even in the neutral case \cite{Faedo:2015jqa}. 
Nevertheless, it turns out that the neutral C-metric does not shield the curvature singularity by the event horizon, implying that the solution fails to describe 
the accelerated black holes with a scalar hair.  Solutions with at least one nonvanishing charge can avoid naked curvature singularities. 
Another insightful outcome of \cite{Nozawa:2022upa} is that the ``flipping transformation'' 
inherent to the C-metric brings the solution into another family of C-metrics found in \cite{Lu:2014ida,Lu:2014sza}, which occurs with the sign change of the scalar field. 
In the case of zero acceleration limit, 
the C-metric in \cite{Lu:2014ida,Lu:2014sza} reduces to the asymptotically AdS,  electrically charged black hole with
a nontrivial scalar field \cite{Anabalon:2012ta,Feng:2013tza}, which does {\it not} admit a parameter range under which the event horizon exists in the spherical and neutral case \cite{Faedo:2015jqa}.

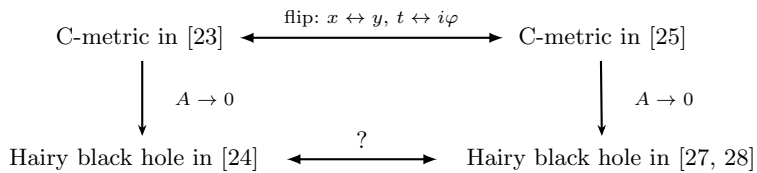
\begin{figure}[ht]
	\begin{center}
		\begin{tikzpicture}
		[
		every node/.style={outer sep=0.15cm, inner sep=0cm},
		arrow/.style={-{Stealth[length=0.15cm]},thick},
		rblock/.style={rectangle, rounded corners,draw, minimum height = 0.5cm,
			minimum width=1.6cm, thick, outer sep = 0},
		point/.style={radius=2pt}
		]
		\node [] (solC){C-metric in \cite{Nozawa:2022upa}\,};
		\node [below=1 of solC] (BH1){Hairy black hole in \cite{Faedo:2015jqa}\,  \, };
		\node [right=3.5 of solC] (flipC){\,C-metric in \cite{Lu:2014ida}};
		\node [right=2 of BH1] (BH2){\hspace{0.4em} Hairy black hole in \cite{Anabalon:2012ta,Feng:2013tza}};
		\draw[latex-latex,thick](solC) -- (flipC) node[above,midway] {{\scriptsize flip:~$x\leftrightarrow y$, $t\leftrightarrow  i\varphi$}};
		\draw[arrow] (solC) -- (BH1) node[right,midway] {{\scriptsize ~~~$A\to 0$}};
		\draw[latex-latex,thick](BH1) -- (BH2) node[above,midway] {{?}};
		\draw[arrow] (flipC) -- (BH2) node[right,midway] {{\scriptsize ~~~$A\to 0$}};
		\end{tikzpicture}
		\caption{Flipping transformation of C-metrics and hairy black holes in supergravity}
		\label{fig:Cmetrics}
	\end{center}
\end{figure}

It should be noted that these two families of hairy solutions are not related by electromagnetic duality and  this flipping transformation is invisible in the zero acceleration limit. Before the discovery of the C-metric in gauged supergravity, 
it has been unclear  why a particular Einstein-scalar system 
gives rise to the two kinds of hairy black holes \cite{Nozawa:2022upa} and \cite{Anabalon:2012ta,Feng:2013tza}. 
With the benefit of hindsight,  the question regarding the appearance of two distinct 
families of static solutions  is a direct consequence of the existence of C-metric and its flipping degrees of freedom. 
This status is schematically shown in figure \ref{fig:Cmetrics}.

With the same objective in mind, 
we consider in this paper the accelerated generalization of wormholes, rather than black holes.
Wormholes describe the nonsingular geometry which allows for tunneling into a different universe through a bridge structure called throat. 
The first physical discussion of traversable wormholes was due to the landmark result of Morris and Thorne in 1988 \cite{Morris:1988cz,Morris:1988tu}. 
They analyzed a static solution with a massless phantom scalar field, which has been later recognized as the same solution previously found by Ellis \cite{Ellis1973} and Bronnikov \cite{Bronnikov1973}. Wormholes offer a theoretical way to perform interstellar and time travels.  The necessity of exotic matter violating energy conditions for the construction of traversable wormholes is an immediate corollary of the topological censorship theorem~\cite{Friedman:1993ty,Galloway}. 
The Ellis-Bronnikov class of wormholes has thus been studied extensively over the years, including studies on 
stability \cite{Nandi:2016ccg,Cremona:2018wkj,Azad:2022qqn}, gravitational lensing \cite{Abe:2010ap,Yoo:2013cia,Tsukamoto:2016qro}, charged and higher dimensional generalizations \cite{Torii:2013xba,Huang:2019arj,Martinez:2020hjm,Nozawa:2020wet}. 
Recently, wormholes in AdS have garnered the significant attention in the context of quantum entanglement and 
teleportation in gauge/gravity duality \cite{Maldacena:2004rf,Gao:2016bin,Maldacena:2017axo,Maldacena:2018lmt}.

In \cite{Nozawa:2020gzz} (see also \cite{Huang:2020qmn}), one of the present authors has  successfully constructed two distinct families of AdS wormhole solutions 
in Einstein-phantom scalar system with a potential. 
These wormhole solutions interpolate the two nearby AdS local minima of the potential. 
A striking property of two families of wormhole solutions is that 
they admit a scalar field  profile with opposite signs. It is then reasonable to hope that these solutions are likely to be related by means of the 
flipping symmetry of a more general C-metric solution, similar to the case of black holes. This expectation is indeed true, as we will demonstrate in this paper.

We present a new C-metric solution in Einstein's gravity with a phantom scalar field. The scalar field has a potential written in terms of 
the superpotential, which is obtainable by a suitable Wick rotation of parameters in ${\cal N}=2$ gauged supergravity model
considered in \cite{Nozawa:2022upa,Faedo:2015jqa,Lu:2014ida,Lu:2014sza,Anabalon:2012ta,Feng:2013tza}. 
The potential admits countably many AdS extrema among which the local minima correspond to the critical points of the superpotential. 
The flipping transformation of the solution brings the present metric into a different family of C-metrics. 
By taking the zero acceleration limit, each solution recovers two wormhole solutions in  \cite{Nozawa:2020gzz}, 
as anticipated. 
We confirm that the instruction presented  in figure \ref{fig:Cmetrics} has proven to be universal also for wormholes. 
Within a natural coordinate domain, our C-metric asymptotes to AdS at the origin of the potential. By the maximal extension, the solution is 
patched smoothly into the other side of the universe,  where the scalar field evolves towards a different critical point of the potential. 
This property mirrors the behavior of a soliton which interpolates two different vacua. 
A noteworthy property of our C-metrics is that for appropriate parameters they do not necessitate conical singularities 
to induce acceleration, for which the acceleration of wormholes is supplied solely by the phantom field.

The present paper is structured as follows. 
In the next section, we provide a detailed description of our model for the Einstein-phantom scalar gravity. 
In section \ref{sec:sol}, a new C-metric solution is presented. We determine the 
physical meaning of parameters of the solution by considering suitable limits. Of particular interest 
to be highlighted is that the C-metric is endowed with the flipping symmetry. 
Physical properties of the C-metric is explored in section \ref{sec:phys}. 
Section \ref{sec:PD} spells out the global causal structure of our solution by drawing 
Penrose diagrams. We summarize our work in section \ref{sec:summary}. 
An appendix presents an alternative C-metric solution with a different superpotential, which 
is the accelerated generalization of the Ellis-Gibbons class of metrics. 

Our conventions of curvature tensors are 
$[\nabla _\rho ,\nabla_\sigma]V^\mu ={R^\mu }_{\nu\rho\sigma}V^\nu$ 
and ${R}_{\mu \nu }={R^\rho }_{\mu \rho \nu }$.
The Lorentzian metric is taken to be the mostly plus sign, and 
Greek indices run over all spacetime indices. 
To maintain simplicity of equations, we work in units $8\pi G=c=1$.

\section{Einstein's gravity with a phantom scalar field}
\label{sec:system}

Let us consider the four dimensional Einstein's gravity with a real scalar field described by Lagrangian
\begin{align}
\label{Lag}
\ma L=R -\epsilon (\nabla\phi)^2 -2 \ma V(\phi)\,,
\end{align}
where $\epsilon=-1$ corresponds to the phantom field. We maintain $\epsilon$ throughout the paper 
to emphasize the consequence of phantom property. 
We consider a theory in which the potential $\ma V(\phi)$ of the phantom field is expressed in terms of a subsidiary function $\ma W(\phi)$ as 
\begin{align}
\label{pot:superpotential}
\ma V(\phi)=4\left[2\epsilon \left(\frac{\partial}{\partial \phi}\ma W(\phi)\right)^2-3 \ma W(\phi)^2 \right]\,. 
\end{align}
By a slight abuse of terminology, we refer to $\ma W(\phi)$ as 
the ``superpotential'' in this paper,  following the language of supergravity ($\epsilon=1$). 
We focus on the theory whose superpotential is given by 
\begin{align}
\label{W}
\ma W(\phi)=\frac{g}{2} e^{\frac{\beta}{\sqrt{{2(1+\beta^2)}}} \phi}\left[
\cos \left(\frac{\phi}{\sqrt{{2(1+\beta^2)}}} \right)-\beta \sin\left(
\frac{\phi}{\sqrt{{2(1+\beta^2)}}} \right)\right] \,,
\end{align}
where $\beta $ is a parameter governing the decay of the potential and $g$ determines the overall scale.  
By the redefinition $\ma W\to -\ma W$ if necessary, one can assume $g\ge 0$ without losing any generality. 
This superpotential makes contact with the ${\cal N}=2$ supergravity model
considered in \cite{Nozawa:2022upa,Faedo:2015jqa,Lu:2014ida,Lu:2014sza,Anabalon:2012ta,Feng:2013tza} by analytic continuation
$\phi\to i\phi/\sqrt 2$ and $\beta\to -i(1-\alpha^2)/(1+\alpha^2)$ with $\alpha \in \mathbb R$.

\begin{figure}[t]
\begin{center}
\includegraphics[width=13cm]{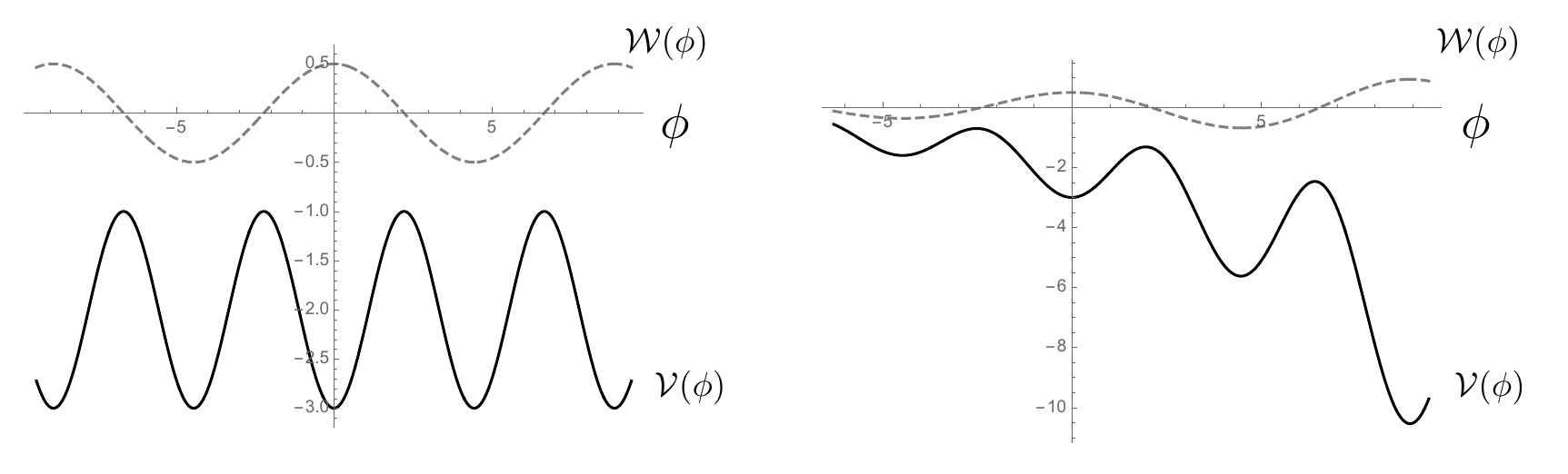}
\caption{Potential $\ma V$ (thick line) and superpotential $\ma W$ (dashed line) for
$\beta=0$ (left figure) and $\beta=0.1$ (right figure). }
\label{fig:pot}
\end{center}
\end{figure}

Typical behaviors of the potential and the superpotential are depicted 
in figure~\ref{fig:pot}. 
For $\beta =0$, both of $\ma V$ and $\ma W$ are  periodic. 
The present potential $\ma V$ admits an infinite number of AdS critical points labeled by 
integer $n\in \mathbb Z$ as 
\begin{align}
\label{cpts}
\phi_n= \sqrt{2(1+\beta^2)}n \pi \,, \qquad 
\ti \phi_n= \left\{
  \begin{aligned}
  & \sqrt{2(1+\beta^2)}\left[n \pi+\arctan\left(\frac 1{2\beta }\right)\right]
 \qquad &(\beta\ne 0)\,,
  \\
  & \sqrt{2}\left(n+ \frac12\right)\pi 
  \qquad &(\beta= 0)\,.
  \end{aligned}
\right.  
\end{align}
The local minima $\phi_n$ of the potential $\ma V$ also extremize the superpotential $\ma W$, which 
is reminiscent of supersymmetric vacua in supergravity. 
The local maxima $\ti \phi_n$ of the potential $\ma V$, on the other hand,  do not correspond to the 
critical points of the superpotential $\ma W$.

At $\phi=\phi_n$, we have\footnote{The mass eigenvalue should have $\epsilon$ term, 
which can be recognized by multi-field covariant expression
$(m^2)^I{}_J=G^{IK}\partial_K\partial_J V$ for $\ma L=-\frac 12 G_{IJ}\partial \phi^I \partial \phi^J-V(\phi^K)$.}
\begin{align}
\label{AdSradii}
\ma V(\phi_n)=-3 g_n^2 \,, \qquad g_n\equiv  ge^{n\pi \beta} \,, \qquad 
m^2_n\equiv \epsilon \ma V''(\phi_n)=-2g_n^2 \,. 
\end{align}
It follows that $g_n$ corresponds to the reciprocal of each AdS radius at $\phi_n$
and the (super)potential is invariant under $\phi\to \phi+\phi_n$ with $g\to g_n$.  
At $\phi=\ti \phi_n$, we obtain
\begin{align}
\label{}
\ma V(\ti \phi_n)=-3 \ti g_n^2 \,, \qquad 
\ti g_n\equiv  \frac{g}{\sqrt 3}e^{\beta(n\pi +\arctan(1/(2\beta))}
 \,, \qquad 
\ti m^2_n\equiv \epsilon \ma V''(\ti \phi_n)=6 \ti g_n^2 \,. 
\end{align}

Note that the construction of exact solutions for $\ma V\ne 0$ is a challenging task to implement. 
A principal adversity in this regard  is that the solution generating technique developed for the 
asymptotically flat case is inoperative, due to 
the presence of the scalar potential \cite{Klemm:2015uba}. 
This fact has thus hampered the systematic analysis
and classification of asymptotically AdS solutions. 
We would like to emphasize that the construction of exact solutions to the present system
is invaluable in its own right.

\section{Wormhole C-metric}
\label{sec:sol}

In this section, we present a new solution corresponding to the accelerated wormholes. 
We clarify the physical meaning of the parameters of the solution by considering the appropriate limits. 
We highlight the existence of flipping transformation of the C-metric, which 
enables us to obtain two distinct families of AdS wormholes in \cite{Nozawa:2020gzz}.

\subsection{Solution}

A new C-metric solution for the system (\ref{Lag}), (\ref{pot:superpotential}) and (\ref{W}) reads\footnote{
\label{fn:construction}
Even though there is no systematic way to derive the 
solution for the $g\ne 0$ case, the present solution (\ref{sol}) has been constructed  by demanding the following reasonable ansatz. (i) The scalar field
should be independent of the AdS radius $g$. (ii) Writing $y=-1/(Ar)$ and $x=\cos\theta$ in the 
static wormholes in \cite{Nozawa:2020gzz}, the solution should admit the flipping symmetry (\ref{flip}). 
(iii) In the $g=0$ case, $(x-y)^4\rho^2$ in (\ref{rhoz}) should be a product of quartic functions of $x$ and $y$, 
as for the Weyl metric. These conditions may be used to fix the metric's dependence on the functions $U$ and $V$ almost uniquely. 
Finally, the structure functions ($\Delta_x, \Delta_y$) are determined by enforcing Einstein's equations. 
}
\begin{subequations}
\begin{align}
\label{}
\D s^2 =& \frac{1}{A^2(x-y)^2}\left[
e^{2\beta U(x)}V(x) \left(-e^{-2\beta U(y)}\Delta_y(y)\D t^2 
+\frac{\D y^2}{e^{-2\beta U(y)}\Delta_y (y)}\right)\right.\notag \\
&\left. 
+e^{2\beta U(y)}V(y)\left(e^{-2\beta U(x)}\Delta_x(x)\D \varphi^2 
+\frac{\D x^2}{e^{-2\beta U(x)}\Delta_x (x)}\right)
\right] \,, \\
\phi =&\sqrt{-2\epsilon(1+\beta^2)}\left(U(x)-U(y)\right)\,.
\end{align}
\label{sol}
\end{subequations}
The reality of the scalar field requires $\epsilon=-1$, i.e., the phantom field. 
The metric comprises four functions ($U, V, \Delta_x, \Delta_y$) given by
\begin{align}
\label{UV}
U(y)=-\arctan\left(\frac{a_1+2 a_2 y}{\sqrt{4a_0a_2-a_1^2}}\right)\,, \qquad 
V(y)=a_0+a_1 y+a_2 y^2 \,, 
\end{align}
and 
\begin{align}
\label{Deltaxy}
\Delta_y(y)=c_0+c_1y+c_2 y^2 \,, \qquad 
\Delta_x(x)=-c_0-c_1x-c_2 x^2 +\frac{g^2}{A^2}V(x) e^{4\beta U(x)}\,, 
\end{align}
Apart from $g$ and $\beta$, the solution possesses seven parameters  
$a_{0,1,2}$, $c_{0,1,2}$ and $A$. 
We see that the scalar field configuration given by (\ref{sol}) tends to vanish at $x=y$. 
As will be discussed in the next section, $x=y$ corresponds to AdS infinity. 
For the spacetime with the wormhole, there come out two types of AdS infinity. The first one occurs when $y$ approaches $x$ from below, i.e., $x = y - $. This AdS infinity for the current solution corresponds to the origin $\phi=0$ of the scalar field potential.\footnote{
One can also consider the configuration around any local minima $\phi_n$ of the potential by 
$\phi\to \phi+\phi_n$ in (\ref{sol}), which can be compensated by $g\to g_n$ in $\Delta_x$. 
By using this freedom, we can assume that $\phi$ tends to zero at AdS infinity, where $x = y - $.
} 
At the other AdS infinity, where $x = y + $, the scalar field approaches one of the two neighboring vacua $\phi_{\pm 1}$
of the superpotential, as we will illustrate in section \ref{sec:PD}.

The solution (\ref{sol}) admits mutually commuting hypersurface-orthogonal Killing vectors
$\partial/\partial t$ and $\partial/\partial \varphi$, and falls into Petrov type D. 
These properties are common to the ordinary C-metric in Einstein-Maxwell-$\Lambda$
 system.

The solution allows the following 
shift and scaling symmetry \cite{Hong:2003gx}
\begin{align}
\label{shift}
x=b_0 x'+b_1\,, \quad 
y=b_0y'+b_1\,, \quad 
t=b_2 t'\,, \quad 
\varphi=b_2 \varphi' \,,
\end{align}
with $A'=\pm \sqrt{b_0b_3/b_2}A$ and 
\begin{align}
\label{}
a_0'=&\,b_3(a_0+b_1a_1+b_1^2a_2) \,, &
a_1'=&\,b_3b_0 (a_1+2b_1a_2)\,, &
a_2'=&\,b_3b_0^2 a_2 \,, \notag \\
c_0'=&\,\frac{b_2}{b_0}(c_0+b_1c_1+b_1^2c_2)\,, &
c_1'=&\,b_2(c_1+2b_1c_2)\,, &
c_2'=&\,b_0b_2 c_2 \,. 
\end{align}
Here, $b_0 (>0)$, $b_1$, $b_2 (>0)$ and $b_3(>0)$ are constants. 
A tailor-made choice of the sign 
allows us to set $A>0$, which will be assumed in the sequel.  
Supposed $a_2\ne 0$, one can always choose $a_1$ to vanish. 

The solution (\ref{sol}) does not permit the $A\to 0$ limit in an obvious fashion, 
because of the overall factor $A^{-2}$. 
To evade this adversity, we exploit the following rescaled coordinates
\begin{align}
\label{rtau}
r=-\frac 1{Ay}\,, \qquad 
\tau =\frac 1A t \,,
\end{align}
in terms of which the solution is rewritten into 
\begin{subequations}
\begin{align}
\label{}
\D s^2 =& \frac{1}{(1+A r x)^2}\left[
e^{2\beta U(x)}V(x) \left(-e^{-2\beta U_r(r)}\Delta_r(r)\D \tau^2 
+\frac{\D r^2}{e^{-2\beta U_r(r)}\Delta_r (r)}\right)\right.\notag \\
&\left. 
+r^2 e^{2\beta U_r(r)}V_r(r)\left(e^{-2\beta U(x)}\Delta_x(x)\D \varphi^2 
+\frac{\D x^2}{e^{-2\beta U(x)}\Delta_x (x)}\right)
\right] \,, 
\\
\phi=& \sqrt{-2\epsilon(1+\beta^2)}\left(U(x)-U_r(r)\right)\,,
\end{align}
\label{rxsol}
\end{subequations}
where
\begin{align}
\label{}
U_r(r)=\arctan \left(\frac{a_2}{\sqrt{a_0a_2}Ar}\right)\,, \qquad 
V_r (r)=a_0 +\frac{a_2}{A^2r^2} \,, \qquad 
\Delta_r(r)=c_2-A c_1 r+A^2 c_0 r^2 \,. 
\end{align}
In the above we have used (\ref{shift}) to set $a_1=0$. 
To keep the Lorentzian signature, we need 
$V(x)>0$ and $V_r(r)>0$, requiring $a_0 $ and $a_2$ to be positive.  
By tuning $b_3$, we can fix $a_0=1$.  
Writing $a_2=m^2 A^2 (>0)$, the $b_2$ degree of freedom in (\ref{shift}) can be used to set 
$k\equiv c_2-g^2 m^2(1+8\beta^2)$ as $k=\{0, \pm 1\}$. 
The remaining parameters $c_0$ and $c_1$
can be fixed by requiring the $A\to 0$ limit to be well-defined, giving rise to 
\begin{align}
\label{}
c_0=\frac{g^2}{A^2}-1\,, \qquad 
c_1=-\frac{4g^2 m\beta }{A} \,,
\end{align}
where $O(A^0)$ term of $c_0$ has been set to be $-1$ by adjusting $b_0$.
It is critical for the present parametrization that the sign change $m\to -m$ can be offset by $\beta\to -\beta$. 
This symmetry allows us to set $m\ge 0$ with no loss of generality.\footnote{
When the theory parameter $\beta$ is fixed, there appear solutions with $m \geq 0$ and $m < 0$, which discriminates 
the physical quantity of the spacetime.  Therefore, it appears more reasonable from physical perspective to confine ourselves to the $\beta \ge 0$ case by varying $m$. 
However, this choice compels us to encounter a serious difficulty when we try to explore the global extension of spacetime, since the asymptotic values of 
 $U(y)$ and $U_r(r)$ at the coordinate boundary depend sensitively on the sign of $m$. To streamline the analysis, we limit the range of $m$ to non-negative values, while  obtaining an equivalent solution with negative $m$ by using $\beta \to -\beta $.
}

To sum up, metric functions are determined to be 
\begin{subequations}
\begin{align}
V(x)=& \,1+A^2m^2 x^2 \,, \\
U(x)=&\,-\arctan (Am x)\,,  \\
\Delta_x(x)=&\, 1-k x^2+\frac{g^2}{A^2}\left(-1+4Am\beta x-A^2 m^2(1+8\beta^2)x^2+e^{4\beta U(x)}(1+A^2m^2 x^2)\right)\,, 
\end{align}
\label{fun1}
\end{subequations}
and
\begin{subequations}
\label{fun2}
\begin{align}
\Delta_r(r)=&\, k+(1+8\beta^2)m^2g^2+4g^2m \beta r+(g^2-A^2)r^2 \,,\\
U_r(r)=&\, \left\{ 
\begin{array}{cc}
\dfrac{\pi}{2}-\arctan\left(\dfrac{r}{m}\right) & (m>0)\,, \\
0 &  (m=0)\,, 
\end{array}
\right.
\\
V_r(r)=&\,   1+\frac{m^2}{r^2}\,.
\end{align}
\end{subequations}
Going back to the $y$ coordinate, 
\begin{align}
\label{fun3}
\Delta_y(y)=-1+\frac{g^2}{A^2}-\frac{4g^2 m\beta y}{A}+\left(k+(1+8\beta^2)m^2g^2\right)y^2\,.
\end{align}
For $g=0$, $\Delta_x(x)=1-kx^2$. Thus, $k$ controls the topology of $x$-$\varphi$ surface: 
$S^2$ for $k=1$, $\mathbb R^2$ for $k=0$ and $\mathbb H^2$ for $k=-1$. 
It turns out that the solution is specified by three physical parameters $m$, $A$ and $k$ 
together with  theoretical parameters  $g$ and $\beta$. The physical meaning of other parameters will be clarified soon. 

It should be emphasized that 
we have derived the expression of $U_r(r)$ in (\ref{fun2}) 
by making use of the formula $\arctan(z)=\pi/2-\arctan(1/z)$ for $z>0$. 
With the current expression of $U_r(r)$, it is evident that $r=0$ is not the 
coordinate singularity,  and the solution (\ref{rxsol}) becomes smooth at $r=0$ ($y\to -\infty$).
If $U_r(r)$ were defined as $U_r(r)=\arctan(m/r)$, the solution would not be
smooth at $r=0$. This substitution  plays a crucial role when considering the maximal
analytic extension of the spacetime, as will be argued in section \ref{sec:PD}.

\subsection{$m=0$ case: AdS}

Setting $m=0$ for the solution (\ref{sol}) with (\ref{fun1}) and (\ref{fun3}), the scalar field vanishes and the metric reduces to 
\begin{align}
\label{AdS}
\D s^2=\frac 1{A^2(x-y)^2}\left(-\Delta_y(y)\D t^2+\frac{\D y^2}{\Delta_y(y)}+\D \Sigma_k^2 (x, \varphi)\right)\,, 
\end{align}
where $\Delta_y(y)=g^2A^{-2}-1+k y^2$ and
\begin{align}
\label{}
\D \Sigma_k^2 (x, \varphi)\equiv \frac{\D x^2}{1-kx^2}+(1-k x^2)\D \varphi^2\,.  
\end{align}
$\D \Sigma_k^2 (x, \varphi)$ denotes the two dimensional metric 
of constant Gauss curvature $k=0, \pm 1$. 
The Riemann tensor for the metric (\ref{AdS}) is simplified to $R_{\mu\nu\rho\sigma}=-2g^2g_{\mu[\rho}g_{\sigma]\nu}$, implying that 
that the spacetime is AdS. 
To see this explicitly, 
we suppose  temporarily $g^2>A^2$ and define new coordinates
\begin{align}
\label{corrdAdS}
R=\frac{\sqrt{F_0(x, y)}}{\sqrt{g^2-A^2}gA(x-y)}\,, \qquad 
w=\frac{g^2x-A^2(x-y)}{\sqrt{F_0(x,y)}}\,, \qquad 
T=\frac{\sqrt{g^2-A^2}}{Ag}t\,, 
\end{align}
where $F_0(x,y)\equiv g^2[g^2-A^2(1-k y^2)]-A^2k (g^2-A^2)(x-y)^2$. 
In terms of these coordinates, the metric (\ref{AdS}) is expressed by the standard static coordinates of AdS as
\begin{align}
\label{AdSstatic}
\D s^2= -\left(k+g^2 R^2\right)\D T^2+\frac{\D R^2}{k+g^2 R^2}+R^2 
\D \Sigma_k^2(w, \varphi) \,.
\end{align}
The above coordinate system (\ref{AdS}) is the counterpart  of the Rindler coordinates in Minkowski spacetime. 
To see this,  consider a static observer sitting at $|y|\to \infty$ with constant $x, \varphi$. 
It can be verified that this observer undergoes an acceleration $a^\mu =u^\nu \nabla_\nu u^\mu$ 
with constant magnitude $|a^\mu |=A$, which enables us to identify $A$ as the acceleration parameter. 

Note that the only  way to obtain $\phi=0$ for the solution (\ref{sol}) is to set $m=0$, resulting in 
AdS. Hence, this solution (\ref{sol}) does not embrace the AdS C-metric sourced by 
a pure cosmological constant \cite{Podolsky:2002nk,Dias:2002mi}.

\subsection{$A=0$ case: wormhole in AdS}

Taking the $A\to 0$ limit of (\ref{rxsol}) with (\ref{fun1}) and (\ref{fun2}), we get 
\begin{subequations}
\label{EBAdSm}
\begin{align}
\D s^2=&-e^{-2\beta U_r(r)}\Delta_- (r)\D \tau^2
+V_r(r)e^{2\beta U_r(r)}
\left(\frac{\D r^2}{V_r(r)\Delta_-(r)}+r^2\D \Sigma_k^2 (x, \varphi)\right)\,, 
\\
\phi=&- \sqrt{-2\epsilon (1+\beta^2)}U_r(r) \,, 
\label{EBAdSmphi}
\end{align}
\end{subequations}
where $V_r(r)$ and $U_r(r)$ are given by (\ref{fun2}) and 
\begin{align}
\label{}
\Delta_-(r)=k+g^2 \left(r^2+4m\beta r+(1+8\beta^2)m^2\right)\,. 
\end{align}
In the spherically symmetric case ($k=1$), 
the solution  (\ref{EBAdSm}) reduces to the static wormhole in AdS  \cite{Nozawa:2020gzz}. 
Thus, the solution  (\ref{EBAdSm}) represents its topological generalization. 
Since $\phi=0$ ($m=0$) reduces to AdS, the solution (\ref{EBAdSm}) 
is not connected to AdS Ellis-Bronnikov wormhole sourced by 
a phantom scalar and a pure cosmological constant.
The $k=0$ Ellis-Bronnikov solution in Einstein-$\Lambda$ system with a massless 
phantom scalar has been obtained in \cite{Wu:2022gpm} with the help of AdS/Ricci-flat correspondence \cite{Caldarelli:2012hy}, while 
the spherical solution $k=1$ is  yet-to-be analytically found \cite{Blazquez-Salcedo:2020nsa}.

If we take $g=0$ with $k=1$, the solution (\ref{EBAdSm}) recovers the 
Ellis-Bronnikov wormhole in the asymptotically flat spacetime \cite{Ellis1973,Bronnikov1973}. 
It should be worth commenting that the parameter $\beta$ remains present in the solution (\ref{EBAdSm}) despite being initially introduced in the superpotential $\cal{W}(\phi)$ which vanishes in the limit as $g\to 0$. It turns out that  $\beta $ is demoted to the 
physical parameter from the theoretical parameter in the $g=0$ case.

For the present paper to be self-contained,  
it is enlightening here to explore the asymptotic structure of the solution, 
by repeating the analysis in  \cite{Nozawa:2020gzz}. 
Around $r\to \infty$, the solution can be expanded as 
\begin{align}
\label{}
 \D s^2\simeq &-\left(k-\frac{2M_{r>0}}{R}+g^2 R^2\right)\D \tau^2+\frac{\D R^2}{k+\gamma_{r>0} -2M_{r>0}'/R+g^2 R^2}+R^2 \D \Sigma_k^2(x,\varphi) \,, \notag \\
 \phi\simeq & \frac{\phi_-^{r>0}}{R}+\frac{\phi_+^{r>0}}{R^2} \,, 
\end{align}
where $R=|r|e^{\beta U_r(r)}V_r(r)^{1/2}$ is the areal radius and 
\begin{align}
\label{}
\phi_-^{r>0}=-\sqrt{2(1+\beta^2)}m \,, \qquad 
\phi_+^{r>0}=-\frac{\beta}{\sqrt{2(1+\beta^2)}}(\phi_-^{r>0})^2 \,, \qquad 
\gamma_{r>0}=-\frac{g^2}2(\phi_-^{r>0})^2 \,,
\end{align}
with 
\begin{align}
\label{}
M_{r>0}=m\beta \left(k+\frac{4}{3}g^2 m^2(1+4\beta^2)\right)\,, \qquad 
M'_{r>0}=M_{r>0}+\frac 23 g^2 \phi_-^{r>0}\phi_+^{r>0}\,.
\end{align}
Instead of $M_{r>0}'$, $M_{r>0}$ corresponds to the physical mass of the spacetime
\cite{Hertog:2004dr,Henneaux:2006hk}. It follows that the scalar field obeys the 
Robin boundary conditions at AdS boundary, which have its roots in the range of 
the mass eigenvalue $-(9/4)g^2\le m^2\le -(5/4)g^2$ given in (\ref{AdSradii}) \cite{Ishibashi:2004wx}. 
Specifically, the slower fall-off mode $\phi_-$ of the scalar field gives a backreaction to the geometry and 
is ascribed as the cause of nonstandard fall-off term $\gamma_{r>0}$, compared to 
the ordinary Dirichlet boundary conditions \cite{Henneaux:1985tv,Hollands:2005wt}.

Since none of curvature invariants constructed out of metric (\ref{EBAdSm}) diverge, one can extend the spacetime 
across the $r=0$ surface into the $r<0$ region.\footnote{As we noticed, this conclusion is unattainable 
if we were to define $U_r(r)=\arctan(m/r)$ rather than $U_r(r)=\pi/2-\arctan(r/m)$.} 
Then, around $r\to -\infty$
\begin{align}
\label{}
\D s^2 \simeq &\,-\left(k-\frac{2M_{r<0}}{R}+g^2_{-1} R^2\right)\D \tau_{r<0}^2
+\frac{\D R^2}{k+\gamma_{r<0} -2M_{r<0}'/R+g^2_{-1} R^2}+R^2 \D \Sigma_k^2(x,\varphi) \,, \notag \\
  \phi\simeq & \,\phi_{-1}+\frac{\phi_-^{r<0}}{R}+\frac{\phi_+^{r<0}}{R^2} \,, 
\end{align}
where $\tau_{r<0}=e^{-\pi \beta}\tau$ 
and 
$\phi_{-1}$, $g_{-1}$ are given respectively by (\ref{cpts}) and (\ref{AdSradii}). 
The asymptotic expansion of scalar field reads 
\begin{align}
\label{}
\phi_-^{r<0}=\sqrt{2(1+\beta^2)}m e^{\pi \beta} \,, \qquad 
\phi_+^{r<0}=-\frac{\beta}{\sqrt{2(1+\beta^2)}}(\phi_-^{r<0})^2 \,.
\end{align}
The coefficients of metric expansion are $\gamma_{r<0}=-g^2_{-1}(\phi_-^{r<0})^2/2$ and 
\begin{align}
\label{}
M_{r<0}=- e^{\pi \beta} M_{r>0}\,, \qquad 
M'_{r<0}=M_{r<0}+\frac 23 g_{-1}^2 \phi_-^{r<0}\phi_+^{r<0}\,.
\end{align}

For the metric (\ref{EBAdSm}) to be eligible as a wormhole, the static Killing vector 
$\partial/\partial \tau$ should be globally timelike, i.e., $\Delta_-(r)>0$. 
Without extra effort, one can verify that $\Delta_-(r)$ is always positive for $k=1, 0$. 
For $k=-1$, $\Delta_-(r)>0$ is guaranteed, provided
\begin{align}
\label{}
g^2 m^2 >\frac{1}{1+4\beta^2} \,. 
\end{align}
Under this condition, the solution (\ref{EBAdSm}) is regarded as a regular topological wormhole in AdS. 
Note that the extension to $r<0$ is asymmetric for $\beta \ne 0$, since the mass of each asymptotic AdS region
disagrees $M_{r>0}\ne M_{r<0}$. This asymmetry is encoded also into the locus of the wormhole throat, corresponding to the minimum of the areal radius $R_{\text{throat}} = m\sqrt{1+\beta^2}e^{\beta(\pi/2-\arctan \beta)}$ at $r=m\beta $, which differs from the coordinate boundary $r=0$.  
When the theoretical parameter $\beta$ is fixed, the throat radius is determined by the parameter $m$. Thus, we can consider $m$ as the physical parameter that defines the size of the wormhole throat.
In the case of $\beta = 0$, the wormhole exhibits symmetry with respect to the throat at $r = 0$, and both regions have a vanishing mass.

\subsection{Flipping transformation}

Working with the general metric functions (\ref{UV}), (\ref{Deltaxy}) in (\ref{sol}), let us 
consider the following transformation
\begin{align}
\label{flip}
x=\hat y\,, \qquad 
y=\hat x\,, \qquad 
t=i \hat \varphi \,, \qquad 
\varphi=i\hat t\,, 
\end{align}
which flips the role of $(x,y)$ and $(t, \varphi)$. The solution is recast into 
the following form
\begin{subequations}
\label{sol2}
\begin{align}
\label{}
\D s^2 =& \frac{1}{A^2(\hat x-\hat y)^2}\left[
e^{2\beta U(\hat x)}V(\hat x) \left(-e^{-2\beta U(\hat y)}\Delta_{\hat y}(\hat y)\D \hat t^2 
+\frac{\D\hat  y^2}{e^{-2\beta U(\hat y)}\Delta_{\hat y} (\hat y)}\right)\right.\notag \\
&\left. 
+e^{2\beta U(\hat y)}V(\hat y)\left(e^{-2\beta U(\hat x)}\Delta_{\hat x}(\hat x)\D \hat \varphi^2 
+\frac{\D \hat x^2}{e^{-2\beta U(\hat x)}\Delta_{\hat x} (\hat x)}\right)
\right] \,, \\
\phi =&\sqrt{-2\epsilon(1+\beta^2)}\left(U(\hat y)-U(\hat x) \right)\,,
\end{align}
\end{subequations}
where 
\begin{align}
\label{}
\Delta_{\hat x} (\hat x)=c_0+c_1 \hat x+c_2 \hat x^2 \,, \qquad 
\Delta_{\hat y} (\hat y)=-c_0-c_1 \hat y-c_2 \hat y^2+\frac{g^2}{A^2}e^{4\beta U(\hat y)}V(\hat y)\,.
\end{align}
Functions $U$ and $V$ are still given by (\ref{UV}). 
It can be explicitly verified that the solution (\ref{sol2}) also solves the field equations 
derived from the Lagrangian (\ref{Lag}) with (\ref{pot:superpotential}) and (\ref{W}). 
A striking feature of the transformation (\ref{flip}) is that the 
flipped solution (\ref{sol2}) keeps the form of the original C-metric (\ref{sol}), aside from the explicit 
form of structure functions ($\Delta_x, \Delta_y$) and the sign of the phantom scalar field. 
To take the $A\to 0$ limit, we employ rescaled coordinates $\hat r=-1/(A\hat y)$ and $\hat \tau =\hat t/A$. 
It turns out that with the following choice of parameters 
\begin{align}
 a_0=&\,1 \,, &
 a_1=&\,0 &
 a_2=&\,m^2 A^2\,, \notag \\
c_0=&\,1\,, & c_1=&\,0\,, & c_2=&\,-k \,, 
\label{para2}
\end{align}
i.e., 
\begin{align}
\label{}
\Delta_{\hat x}(\hat x)=1-k \hat x^2 \,, \qquad 
\Delta_{\hat y}(\hat y)=-1+k \hat y^2+\frac{g^2}{A^2}e^{4\beta U(\hat y)}V(\hat y)\,, 
\end{align}
where $U$ and $V$ are given by (\ref{fun1}), 
we can take the $A\to 0$ limit. 
Here, we choose $m \geq 0$, similar to the unflipped case.
The $A=0$ solution reads
\begin{subequations}
\label{EBAdSp}
\begin{align}
\D s^2=&-e^{-2\beta U_{r}(\hat r)}\Delta_{+} (\hat r)\D \hat \tau^2
+V_{r}(\hat r)e^{2\beta U_{r}(\hat r)}
\left(\frac{\D \hat r^2}{V_{r}(\hat r)\Delta_+ (\hat r)}+\hat r^2 \D\Sigma_k^2(\hat x, \hat \varphi)\right)\,, 
\\
\phi=&+ \sqrt{-2\epsilon (1+\beta^2)}U_{ r}(\hat r) \,, 
\label{EBAdSpphi}
\end{align}
\end{subequations}
where $U_r$ and $V_r$ are given by (\ref{fun2}) and 
\begin{align}
\label{}
\Delta_+(\hat r)&\equiv k+g^2 \hat r^2 V_{ r}(\hat r)e^{4\beta U_{r}(\hat r)}\,.
\end{align}
It should be noted that the choice of parameters (\ref{para2}), which is demanded by the existence of the $A\to 0$
limit, differs from the previous one. 

Setting $m=0$, the metric is reduced to AdS at $\phi=0$ as in the previous case. 
Compared with (\ref{EBAdSm}), the solution (\ref{EBAdSp}) has a scalar field of 
opposite sign with a  structure function $\Delta_+$ distinct from $\Delta_-$. 
Nevertheless, this solution also describes a regular wormhole in AdS for $k=1$ \cite{Nozawa:2022upa}.

Taking the asymptotic limit $\hat r\to\infty $ of the solution (\ref{EBAdSp}), we have 
\begin{align}
\label{}
 \D s^2\simeq &-\left(k-\frac{2\hat M_{\hat r>0}}{\hat R}+g^2 \hat R^2\right)\D \hat \tau^2
 +\frac{\D \hat R^2}{k+\hat \gamma_{\hat r>0} -2\hat M_{\hat r>0}'/\hat R+g^2 \hat R^2}+\hat R^2 \D \Sigma_k^2(\hat x, \hat \varphi) \,, \notag \\
 \phi\simeq & \frac{\hat \phi_-^{\hat r>0}}{\hat R}+\frac{\hat \phi_+^{\hat r>0}}{\hat R^2} \,, 
 \label{WHex2}
\end{align}
where $\hat R=|r|e^{\beta \hat U_r(\hat r)}\hat V_r(\hat r)^{1/2}$ and 
\begin{align}
\label{}
\hat \phi_-^{\hat r>0}=\sqrt{2(1+\beta^2)}m \,, \qquad 
\hat \phi_+^{\hat r>0}=\frac{\beta}{\sqrt{2(1+\beta^2)}}\left(\hat \phi_-^{\hat r>0}\right)^2 \,, \qquad 
\hat \gamma_{\hat r>0}=-\frac{g^2}2\left(\hat \phi_-^{\hat r>0}\right)^2 \,,
\end{align}
The mass parameters are given by 
\begin{align}
\label{}
\hat M_{\hat r>0}=km\beta \,,  \qquad 
\hat M'_{\hat r>0}=\hat M_{\hat r>0}+\frac 23 g^2 \hat \phi_-^{\hat r>0}\hat \phi_+^{\hat r>0}\,.
\end{align}
Since this solution (\ref{EBAdSp}) is also free of scalar curvature singularities, 
one can extend the physical region $\hat r\ge 0$ to $\hat r<0$. In the asymptotic $\hat r\to -\infty $ limit, 
the solution is approximated as
\begin{align}
\label{}
\D s^2 \simeq &\,-\left(k-\frac{2\hat M_{\hat r<0}}{\hat R}+g^2_1 \hat R^2\right)\D \hat \tau_{\hat r<0}^2
+\frac{\D \hat R^2}{k+\hat \gamma_{\hat r<0} -2\hat M_{\hat r<0}'/\hat R+g^2_1 \hat R^2}+\hat R^2 \D \Sigma_k^2(\hat x, \hat \varphi) \,, \notag \\
  \phi\simeq &\, \phi_{1}+\frac{\hat \phi_-^{\hat r<0}}{\hat R}+\frac{\hat \phi_+^{\hat r<0}}{\hat R^2} \,, 
\end{align}
where $\hat \tau_{\hat r<0}=e^{-\pi \beta}\hat\tau$. $\phi_1$ and $g_1$ are given by (\ref{cpts}), (\ref{AdSradii}), respectively. 
Other coefficients read
\begin{align}
\label{}
\hat \phi_-^{\hat r<0}=-\sqrt{2(1+\beta^2)}me^{\pi \beta} \,, \qquad 
\hat \phi _+^{\hat r<0}=\frac{\beta}{\sqrt{2(1+\beta^2)}}\left(\hat \phi_-^{\hat r<0}\right)^2 \,,
\end{align}
with $\hat \gamma_{\hat r<0}=-g_{1}^2(\hat \phi_-^{\hat r<0})^2/2$ and 
\begin{align}
\label{}
\hat M_{\hat r<0}=- e^{\pi \beta}\hat M_{\hat r>0} \,, \qquad 
\hat M'_{\hat r<0}=\hat M_{\hat r<0}+\frac 23 g_{1}^2 \hat \phi_-^{\hat r<0}\hat \phi _+^{\hat r<0}\,.
\end{align}
$\Delta_+(\hat r)>0 $ is satisfied for $k=0, 1$, leading to the AdS wormhole geometry. 
$\Delta_+(\hat r)>0 $ is ensured for $k=-1$, provided 
\begin{align}
\label{}
g^2m^2 >\frac{e^{-2\beta(\pi-2\arctan(2\beta))}}{1+4\beta^2}\,.
\end{align}
Under this condition, the solution with $k=-1$ also serves as a static wormhole in AdS. 
For either sign of $k$, the wormhole throat is given by $\hat R_{\rm throat}=m\sqrt{1+\beta^2}e^{\beta(\pi/2-\arctan\beta)}$
at $\hat r=m\beta$. 

Let us comment that the asymptotic value $\phi_{1}$  of the flipped wormhole 
(\ref{EBAdSp}) differs from the one $\phi_{-1}$ for unflipped wormhole (\ref{EBAdSm}). 
Viewed from the origin of the potential, 
these are two neighboring critical points of the superpotential. 
This difference is ascribed to the sign flip of the scalar field [see (\ref{EBAdSmphi}) and (\ref{EBAdSpphi})]. 
Since the discovery of these solutions in \cite{Nozawa:2020gzz}, it has remained unclear 
why these different wormhole solutions exist in the same theory. 
We have explicitly demonstrated above that its geometric origin is attributed to the existence of C-metric 
flipping transformation.

\section{Physical properties}
\label{sec:phys}

In this section we embark on the task of uncovering the physical properties  of the C-metric solution (\ref{sol}). For the sake of clarity, 
we restrict ourselves to the case where  $k=1$ and $m>0$. 

\subsection{Conical singularity}

We consider the case in which $\Delta_x(x)$ given by (\ref{fun1}) admits at least 
two real distinct roots $x_\pm$ 
\begin{align}
\label{}
\Delta_x(x_\pm)=0\,, \qquad \Delta_x(x)>0~~(x_-<x<x_+)\,. 
\end{align}
A possible conical singularity at $x=x_+$ can be avoided, 
provided $\varphi_+=e^{-2\beta U(x_+)}(|\Delta_x'(x_+)|/2)\phi$
has a periodicity $2\pi$. However, the conical singularity 
at $x=x_-$ is generically inevitable. Thus, the solution is viewed as a 
wormhole accelerated by a cosmic string. 

Exceptional cases are $\beta=0$ or $g=0$, for which 
$\Delta_x(x)=1-x^2$. In these special cases, the two dimensional surface
spanned by $x$ and $\varphi$
describes a regular $S^2$ without conical singularities. 
Consequently, we do not need distributional sources to maintain thebacceleration of wormholes. 
This property stands in stark contrast to the vacuum case.

\subsection{Infinity}

Let us consider the ``radial'' null geodesics ($\dot x=\dot \varphi=0$) 
described by 
\begin{align}
\label{}
-e^{-2\beta U(y)}\Delta_y(y)\dot t^2
+\frac{\dot y^2}{e^{-2\beta U(y)}\Delta_y (y)}=0 \,, \qquad 
E=\frac{e^{2\beta(U(x)-U(y))}V(x)\Delta_y(y)}{A^2(x-y)^2}\dot t\,, 
\end{align}
where the dot denotes the derivative with respect to the 
affine parameter $\lambda$ of the null geodesics and 
$E$ is a constant corresponding to the energy. 
Upon integration, the affine parameter is given by 
\begin{align}
\label{}
\lambda =\pm \frac{e^{2\beta U(x)}V(x)}{A^2 E (x-y)}\,. 
\end{align}
It turns out that $x=y$ corresponds to infinity, since an infinite 
amount of affine time is needed for radial null geodesics to arrive at $x=y$.
It therefore turns out that the coordinate $r=-1/(Ay)$  is of no utility to explore the 
global causal structure, since $r\to \infty $ can be reached by a finite 
affine parameter for $x\ne 0$.

The proper distance $s $ along the curve $\{t, x, \varphi\}={\rm const.}$ is also 
a useful quantity to reveal the causal structure 
\begin{align}
\label{}
s=\left|\frac{e^{\beta U(x)}V(x)^{1/2}}{A} \int \frac{e^{\beta U (y)}}{(x-y)\Delta_y (y)^{1/2}}\D y \right|\,.
\end{align}
This allows us to deduce if the given point is a spatial infinity or not. 
Besides $x=y$, 
the proper distance becomes infinitely large if $\Delta_y(y)$ has a double root. 
This is the same as what happens for the degenerate event horizon.

\subsection{Curvature singularities}

The spacetime curvature singularity is identified by the 
divergence of curvature invariants $R_{\mu\nu\rho\sigma}R^{\mu\nu\rho\sigma}$, 
$R_{\mu\nu}R^{\mu\nu}$, $R$ etc. Since all of these expressions are not illuminating, 
we do not show them here. One can nevertheless verify that 
a plausible divergence  comes exclusively from  $V(x)V(y)=0$. 
Since we have required the positivity of $V(x)$ and $V(y)$, 
we conclude that our solution is free of any curvature singularities.

\section{Causal structure}
\label{sec:PD}

We are now going to discuss the global causal structure of the C-metric by 
maximal extension. 
In pursuit of  this aim, we first need to specify the coordinate domain. 
Inspecting (\ref{rtau}), 
we designate  the first coordinate domain (I) to be represented by 
\begin{align}
\label{}
{\rm (I):}~~ x-y \ge 0 \,, \qquad \Delta_x(x)\ge 0 \,. 
\end{align}
At the asymptotic AdS infinity in domain (I), the scalar field is given by 
$\phi^{\rm (I)}|_{x=y}=\sqrt{2(1+\beta^2)}(U(x)-U(y))_{x=y}=0$, 
corresponding to the origin of the potential. 

As $y$ decreases in domain (I), one encounters the coordinate boundary $y=-\infty$. 
As we spelled out in previous section, $y=-\infty$ is a regular surface and 
is not infinity. 
One can then continue $y$ across this surface to  $y=+\infty$ side by $r=-1/(Ay)$. 
Under this extension, the second coordinate domain (II) 
for the other side of the universe is covered by
\begin{align}
\label{}
{\rm (II):}~~x-y\le 0\,, \qquad \Delta_x(x)\ge 0 \,. 
\end{align}


The precise extension can be done as follows. 
Suppose $x<0$. In the domain (I) $-\infty <y\le x$, 
 we have $U_y^{\rm (I)}(y)=\arctan (-Am y)>0$. Since $\arctan (-Am y)$ is not smooth at $|y|=\infty $, 
one must replace 
$\arctan(-Amy)\to \pi/2 +\arctan (1/(Amy))$ to traverse the $|y|=\infty $ surface. 
Thus in domain (II-a) $0\le y<\infty $, one obtains $U_y^{\rm (II-a)}(y)= \pi/2 +\arctan (1/(Amy))$. 
As $y$ decreases, one arrives at $y=0$. Then  $ \pi/2 +\arctan (1/(Amy))$ is not smooth there
and one needs another replacement $ \pi/2 +\arctan (1/(Amy))\to \pi -\arctan (Amy)$ in domain (II-a)
to cross $y=0$ surface. 
Hence in domain (II-b) $x\le y<0$, $U_y^{\rm (II-b)}(y)=\pi -\arctan (Amy)=\pi +U^{\rm (I)}(y)$. 
Note that  $U(x)=-{\rm arctan}(Amx)$  remains untouched during these extensions, 
since we are focusing on the fixed and bounded $x$. 
It follows that the asymptotic value of the scalar field in the domain (II) for $x<0$ reads 
$\phi^{\rm (II)}|_{x=y}=\sqrt{2(1+\beta^2)}(U(x)-\pi-U^{\rm (I)}(y))_{x=y}=\sqrt{2(1+\beta^2)}\times (-\pi)=\phi_{-1}$. 

Let us next suppose $x\ge 0$. 
As in the same reasoning above, we have 
$U_y^{\rm (I)}(y)=\arctan (-Am y)$ in domain (I) $-\infty <y<x$, and 
 $U_y^{\rm (II)}(y)= \pi/2 +\arctan (1/(Amy))$ in domain (II) $0\le x<y<\infty $. 
 In this case, the asymptotic value of the scalar field reads
$\phi^{\rm (II)}|_{x=y}=\sqrt{2(1+\beta^2)}\times [-\pi/2-\arctan (Amx)-\arctan (1/(Amy))]_{x=y}=
\sqrt{2(1+\beta^2)}\times(-\pi)=\phi_{-1}$, as in the $x<0$ case. 
It follows that the solution interpolates two nearby critical points of the superpotential, 
regardless of the angular direction $x$. 
This is precisely the same structure as what we have encountered in the 
non-accelerated wormholes.

Since $x$ is recognized as a directional cosine ($x\sim \cos\theta$) of $S^2$, 
the essential spacetime causal structure is determined by the two dimensional portion
\begin{align}
\label{}
\D s_2^2 =-e^{-2\beta U(y)}\Delta_y(y)\D t^2 
+\frac{\D y^2}{e^{-2\beta U(y)}\Delta_y (y)}
=-e^{-2\beta U(y)}\Delta_y(y)\left(\D t^2-\D y_*^2\right)\,, 
\end{align}
where $y_*=\int (e^{2\beta U(y)}/\Delta_y (y) )\D y$ is analogous to the 
tortoise coordinate. 
It therefore follows that the infinite $y_*$ corresponds to the null surface, while 
the finite $y_*$ corresponds to the timelike (spacelike) surface for $\Delta_y(y)>0$ ($<0$). 
Since $e^{2\beta U(y)}$ is positive and bounded, 
the null surface occurs only at $\Delta_y=0$ corresponding to the Killing horizon.

In the following, 
we shall put particular emphasis on the two simplest cases $g=0$ and $\beta=0$, which are amenable to 
analytic study. In each case, the global causal structure is $x$-dependent, as in the case of ordinary C-metric in vacuum.

\subsection{$g=0$ case}

We begin our discussion with the illuminative case of $g=0$, for which the potential of the scalar field vanishes. 
The solution is viewed as an accelerated generalization of the Ellis-Bronnikov wormholes in the asymptotically 
flat spacetime. 
In this case,  we have $\Delta_x(x)=1-x^2$, $\Delta_y(y)=y^2-1$, giving rise to geometry without conical singularities. 
We see that there exists at least one accelerating horizon at $y=-1$ for fixed $x$ ($-1\le x\le 1$).

\begin{figure}[t]
\begin{center}
\includegraphics[width=16cm]{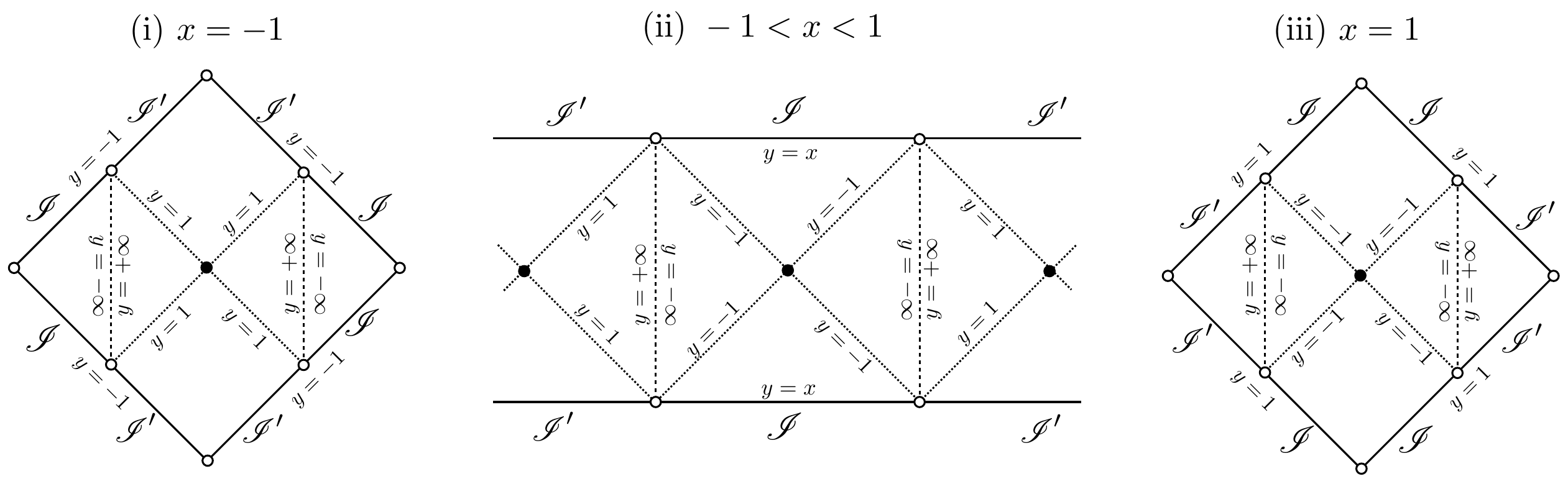}
\caption{Penrose diagrams for $g=0$. 
Dotted lines stand for the Killing horizons, while the dashed lines correspond to the coordinate boundaries $|y|=\infty$. 
The black dots denote the bifurcation surfaces, while white circles denote spatial or timelike infinities. 
The ``scri'' is infinity for null geodesics distinguished by $\mas I$ ($y\to x -$) and $\mas I'$ ($y\to x +$). 
}
\label{fig:PDg0}
\end{center}
\end{figure}

For $x=-1$, the first coordinate domain (I) is $y\le -1$. $y=-1$ therefore corresponds to infinity for null geodesics, which 
possesses the null structure since $y_*$ diverges. 
One can check that  $y\to -\infty$ surface is timelike, 
reached by a finite affine time for radial null geodesics and does not correspond to 
spatial infinity.  One can extend the spacetime across $y=-\infty$ surface 
to $y=+\infty $ side  by $r=-1/(A y)$,  
for which $U(y)=\pi/2+\arctan[1/(Amy)]$ while $U(x)=-\arctan(Am x)$ remains intact. 
As $y$ decreases from $y=+\infty $ in the second coordinate domain (II) $-1\le y$, 
one encounters the null surface $y=+1$, across which the $y<+1$ domain is in the trapped region 
($y={\rm const.}$ surface is spacelike). As $y $ decreases further, one finds the spacelike surface at 
$y=0$, across which the replacement $U(y)=\pi/2+\arctan[1/(Amy)]=\pi/2+[\pi/2-\arctan(Amy)]=\pi-\arctan(Amy)$
is necessary. Further decrement of $y$ reaches null infinity at $y=-1$.
The corresponding Penrose diagram is depicted in (i) of figure~\ref{fig:PDg0}. 

Next, let us consider the case with  $-1<x<1$. For $-1<y\le x$, AdS infinity $y=x $ is in the trapped region and 
has a spacelike structure. As $y$ decreases, one finds the null surface at $y=-1$ and a timelike surface 
$y=-\infty$, which is smooth  and thus extendible to the $y=+\infty$ side. 
After passing $y=+1$, one arrives at infinity $y=x~(-1<x<1)$. The corresponding Penrose diagram is 
(ii) of figure~\ref{fig:PDg0}. 

The Penrose diagram for $x=1$ is deduced similarly.  Infinity at 
$y=1$ lying in the first domain (I) has a null structure. It turns out that the 
global structure is (iii) of figure~\ref{fig:PDg0}, which is essentially the same as (I), 
up to the interchange of coordinate domains (I) and (II).

It is worthwhile to comment that 
the metric in the $g=0$ case falls into the Weyl class 
\begin{align}
\label{}
\D s^2=-e^{2u(\rho, z)}\D t^2+e^{-2u(\rho, z)}\left(\rho^2 \D \varphi^2+e^{2\gamma(\rho, z)}(\D \rho^2+\D z^2)\right)\,, \qquad 
\phi=\phi(\rho, z) \,.
\end{align}
In the $k=1$ case, we have 
\begin{align}
\label{rhoz}
\rho =\frac{\sqrt{(1-x^2)(1+A^2m^2x^2)(y^2-1)(1+A^2m^2y^2)}}{A^2(x-y)^2}
\,, \qquad 
z=\frac{(1-xy)(1+A^2m^2xy)}{A^2(x-y)^2}\,. 
\end{align}
Explicitly, 
the metric components are given by 
\begin{align}
\label{}
u=\beta (U(x)-U(y))+\log\left(\frac{\sqrt{V(x)\Delta_y(y)}}{A(x-y)}\right) \,, \qquad 
e^{2\gamma}=\frac{e^{4\beta U(x)}V(x)^2V(y)\Delta_y(y)}{(1+A^2m^2)^2(y^2-x^2)(1+A^2m^2x^2y^2)} \,.
\end{align}
Functions $u$ and $\phi$ obey axisymmetric Laplace equations on $\mathbb R^3$
\begin{align}
\label{}
\Delta_{\mathbb R^3}u=\Delta_{\mathbb R^3}\phi=0 \,, \qquad 
\Delta_{\mathbb R^3}=\frac{\partial^2}{\partial \rho^2}+\frac{1}{\rho}\frac{\partial}{\partial \rho}+\frac{\partial^2}{\partial z^2}\,.
\end{align}

\subsection{$g\ne 0$ and $\beta=0$ case}

We now turn to the causal structure of  the $g\ne 0$ case. 
First we set $\beta=0$, for which the analytic classification is possible. 
It is also noteworthy that there appear no conical singularities on the axis 
in this distinguished case, for which the source of acceleration is provided by 
a phantom scalar.

Now, it is useful to work with dimensionless quantities
\begin{align}
\label{}
\mathsf A=\frac{A}{g}~(>0)\,, \qquad \mathsf m=mg ~(>0)\,,
\end{align}
in terms of which $\Delta_y(y)=-1+\mathsf A^{-2}+(1+\mathsf m^2)y^2$.
Since  $\Delta_x(x)=1-x^2$, we have symmetry axes at $x=\pm 1$. 

\begin{figure}[t]
\begin{center}
\includegraphics[width=17cm]{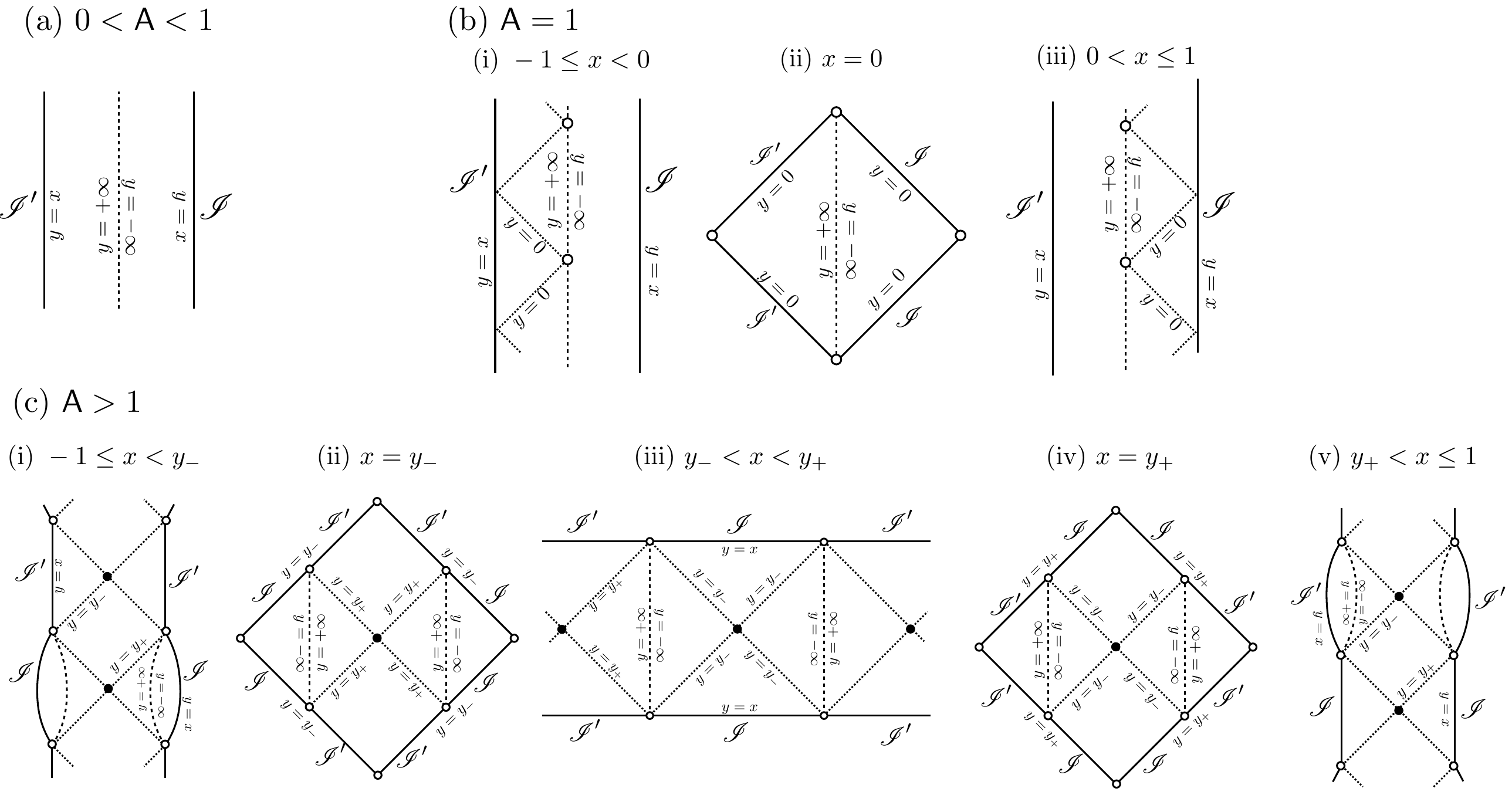}
\caption{Penrose diagrams for $\beta=0$ and $g\ne 0$. Depending on the horizon structure, the diagrams are classified into three subclasses, according to (a) $0<\mathsf A<1$, (b) $\mathsf A=1$ and (c) $\mathsf A>1$.
Dotted lines stand for the Killing horizons, while the dashed lines correspond to the coordinate boundaries $|y|=\infty$. The black dots denote the bifurcation surfaces, while white circles denote spatial or timelike infinities. The ``scri'' is infinity for null geodesics distinguished by $\mas I$ ($y\to x -$) and $\mas I'$ ($y\to x +$).  
}
\label{fig:PDbeta0}
\end{center}
\end{figure}

For $0<\mathsf A<1$, we have $\Delta_y(y)>0$, indicating the absence of a horizon. 
The other asymptotic AdS vacuum at $\phi=\phi_{-1}$ can be glued
 across the timelike $|y|=\infty$ surface to the original vacuum at $\phi=\phi_{0}=0$. 
The global structure is (a) of figure~\ref{fig:PDbeta0} which is 
the same as the static AdS wormhole solutions (\ref{EBAdSm}) and (\ref{EBAdSp}).

For $\mathsf A=1$, the surface $y = 0$ represents a degenerate Killing horizon for $x\ne y$, and 
infinity for $x=y=0$. Away from $y=0$, $\Delta_y(y)>0$ is satisfied. 
The degenerate Killing horizon appears in either coordinate domain (I) or (II).
At $x=0$ there is no horizon, and the global structure corresponds to a configuration that connects two Minkowski-like spacetimes.
The global structure corresponds to (b-i)--(b-iii) of figure~\ref{fig:PDbeta0}.

For $\mathsf A>1$, $\Delta_y(y)=0$ necessarily allows two roots $y=y_\pm$, 
where 
\begin{align}
\label{}
-1<y_-\equiv -\sqrt{\frac{1-\mathsf A^{-2}}{1+\mathsf m^2}}\,, \qquad 
y_+\equiv \sqrt{\frac{1-\mathsf A^{-2}}{1+\mathsf m^2}}<1\,.
\end{align}
These loci correspond to Killing horizons except at $x=y_\pm$.
Depending on the direction $x$, these horizons appear in  each domain.
The global structure is (c-i)--(c-v) of figure~\ref{fig:PDbeta0}. 
Specifically, diagrams (cii), (ciii), and (civ) are the same as diagrams (i), (ii), and (iii) in figure \ref{fig:PDg0}, respectively. However, diagrams 
(c-i) and (c-v) exhibit characteristic features specific to the asymptotically AdS case.

\subsection{$g\beta\ne 0$ case}

Next, let us investigate the most general case $g\beta\ne 0$. 
Unfortunately, the non-polynomial character of $\Delta_x(x)$ given in (\ref{fun1})
reaches a  level of substantial complexity, which makes the exhaustive classification 
and general analysis difficult. Instead of dwelling on this task, we just content ourselves with 
demonstrating that there appears a parameter region under which the wormhole structure 
remains valid also in this case.

\begin{figure}[t]
\begin{center}
\includegraphics[width=14cm]{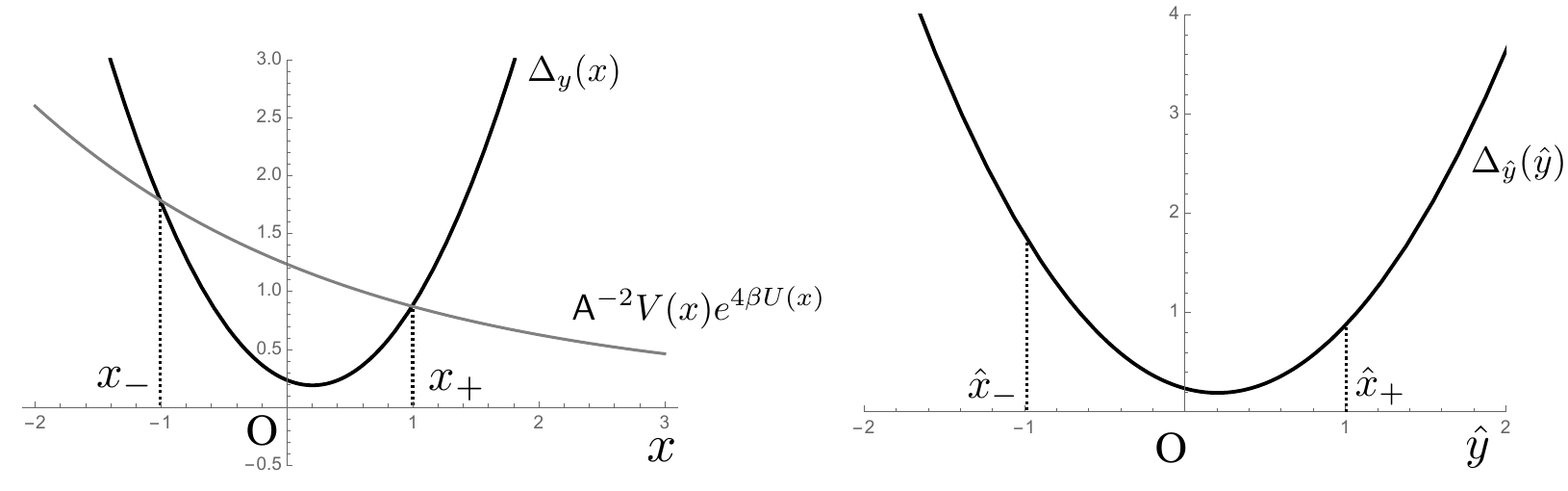}
\caption{Left figure is the plots of $\Delta_y (x)$ (thick line) and $\mathsf A^{-2}V(x) e^{4\beta U(x)}$ (thin line) for 
$\mathsf A=0.9$, $\mathsf m=0.1$ and $\beta =1$. Two intersecting points of these curves 
correspond to the symmetry axis $x_\pm$. 
The right figure represents $\Delta_{\hat y}(\hat y)$ for the same parameter value.
The corresponding plots for changing the sign of $\beta$ are obtained by the reflection of the plot across the vertical axis.}
\label{fig:Deltax}
\end{center}
\end{figure}

As in the case of  supergravity C-metric \cite{Nozawa:2022upa}, 
we find the following notable relation
\begin{align}
\label{univrel}
\Delta_x(x) =\mathsf A^{-2} V(x) e^{4\beta U(x)}-\Delta_y (x) \,,
\end{align}
where $\Delta_y(y)=-1+\mathsf A^{-2}-4 \mathsf m \mathsf A^{-1}\beta y+(1+\mathsf m^2(1+8\beta^2))y^2$, 
$U(x)=-\arctan(\mathsf A \mathsf m x)$ and $V(x)=1+\mathsf A^2\mathsf m^2 x^2$. 
The Killing horizon and the symmetry axis appear respectively at $\Delta_y(y)=0$ and 
$\Delta_x(x)=0$. Using the aforementioned relation (\ref{univrel}), one can visually recognize 
their positional relationship by the intersection of curves $\Delta_y (x)$ or $\Delta_x (x)$ and 
$\mathsf A^{-2}V(x) e^{4\beta U(x)}$. See the left plot in figure \ref{fig:Deltax}. 

Requiring the Killing vector $\partial/\partial t$ is globally timelike, 
the dimensionless acceleration parameter obeys
\begin{align}
\label{}
0<\mathsf A <\sqrt{\frac{1+\mathsf m^2+4\mathsf m^2\beta^2}{1+\mathsf m^2+8\mathsf m^2\beta^2}}<1 \,. 
\end{align}
If, in addition, the $\Delta_x(x)=0$ admits two roots corresponding to the axis, the solution can be 
interpreted as a wormhole. The parameterization shown in figure \ref{fig:Deltax} corresponds precisely 
to this case. The corresponding global structure is the same as (a) in figure \ref{fig:PDbeta0}. 
We have also confirmed that the equation $\Delta_y(x) = 0$ has either one multiple root or two roots, which correspond to the degenerate horizon and two Killing horizons, respectively.
Furthermore, we have found parameter values for which $\Delta_x(x) = 0$ has no real roots, indicating that the spacetime does not exhibit a wormhole structure.

\subsection{Flipped solution}

Finally, let us demonstrate the causal structure for the flipped solution (\ref{sol2}). 
For $g\beta=0$ case, the metric is the same as unflipped one (\ref{sol}).
The difference arises only from the $g\beta \ne 0$ case. 

Extension of the spacetime can be done in a parallel fashion as in the previous case. 
In domain (I) $-\infty< \hat y\le \hat x$, $U^{(\rm I)}(\hat y)= \arctan(-Am\hat y)$, 
so that the scalar field is sitting at the origin of the potential at asymptotic infinity. 
As one crosses $|y|=\infty$ surface, one needs $\arctan(-Am\hat y)\to \pi/2 -\arctan(1/(-Am\hat y))$. 
Thus, in domain (II) $\hat x\le \hat y<\infty$, $U^{(\rm II)}(\hat y)= \pi/2 +\arctan(1/(Am\hat y))$. 
Hence, the asymptotic value of the scalar field in domain (II) is 
$\phi^{(\rm II)}|_{\hat x=\hat y}=\sqrt{2(1+\beta^2)}\times \pi =\phi_1$ for $0\le \hat x$. 
The analysis for $\hat x<0$ case is deduced in the same vein.

We suppose $k=1$ and $m>0$. Then, we obtain $\Delta_{\hat x}(\hat x)=1-\hat x^2$, implying
that $\hat x=\pm 1$ is the symmetry axis. An example of the plot is shown in the right of 
figure \ref{fig:Deltax}. In this case, $\Delta_{\hat y}(\hat y)=-1+\hat y^2+\mathsf A^{-2}e^{4\beta U(\hat y)}V(\hat y)>0$ is satisfied, 
signifying that the global structure is given by (a) in figure \ref{fig:PDbeta0}. 
We have also confirmed the existence of a solution that exhibits a wormhole structure with horizons.

\section{Summary}
\label{sec:summary}

We have constructed a new C-metric solution with a wormhole structure in the Einstein-phantom-scalar system. In the case of a massless scalar field, the solution corresponds to an accelerated generalization of Ellis-Bronnikov wormholes. The scalar potential is  built out of a superpotential and admits an infinite number of extrema. Our model (\ref{W}) can be obtained through the analytical continuation of the parameters in the ${\cal N}=2$ supergravity model. When traversing through the wormhole throat from one universe to another, the scalar field evolves from the origin to the adjacent AdS extrema of the superpotential. This property is redolent of solitons and domain walls.

One of the major advantages of the C-metric is the manifestation of the flipping transformation (\ref{flip}), which allows the solution to be transformed into another C-metric. In the limit of zero acceleration, these solutions revert to the two families of AdS wormhole solutions found in \cite{Nozawa:2020gzz}. 
The same thing happens also for the Ellis-Gibbons class of solutions, as demonstrated in appendix. 
Importantly, the existence of two families of black hole/wormhole solutions is intrinsic to four dimensions \cite{Nozawa:2020gzz}. 
Our conclusion is persuasive and consistent with the absence of the C-metric in higher dimensions \cite{Kodama:2008wf}.

We have also provided a detailed clarification of the global causal structure. The corresponding Penrose diagrams are found in figures \ref{fig:PDg0} and \ref{fig:PDbeta0}. Remarkably, the solution may be free of conical singularities, which is a characteristic not observed in the ordinary C-metric. 
 
Our new solution would offer a window to examine  the spacetime structure and physical properties of wormholes in greater detail. 
A rotating variant of the  C-metric is known as the Pleba\'nski-Demia\'nski solution \cite{Plebanski:1976gy}, which is the 
most general solution in Einstein-Maxwell-$\Lambda$ system falling into Petrov-D type. Physical properties of this solution 
have been discussed in \cite{Griffiths:2005qp,Klemm:2013eca}. 
Although some rotating generalization of Ellis-Bronnikov wormholes have been constructed, e.g., in \cite{Chew:2019lsa,Deligianni:2021hwt}, 
they do not seem to fall into this category (see also \cite{Clement:2022pjr,Barrientos:2023tqb,Cisterna:2023uqf} for recent related works). 
It seems a useful strategy to look for exact solutions from algebraic point of view of Weyl curvature. 

The C-metric admits a shear-free null geodesic congruence without twist. This means that it falls within a Robinson-Trautman class \cite{RT,RT2}. 
We can indeed find a family of Robinson-Trautman solutions for the superpotential (\ref{W}) with $\beta =0$. 
This solution is dynamical and belongs to Petrov type II. We will provide a detailed report on these findings in the near future \cite{NTRT}.

The C-metric solution in Euclidean signature  is also an alluring subject to be examined.
The C-metric instanton solution represents a pair production of black holes by the cosmic string
\cite{Dowker:1993bt,Hawking:1995zn,Eardley:1995au}. 
The Euclidean Pleba\'nski-Demina\'nski solution allows an abundance of mathematically 
rich properties such as the conformal ambi-K\"ahler structure \cite{Nozawa:2015qea,Nozawa:2017yfl} and 
hidden symmetry \cite{Houri:2014hma}.  
Pursuing these issues is  left for future investigation.

\subsection*{Acknowledgements}
The work of MN is partially supported by MEXT KAKENHI Grant-in-Aid for Transformative Research Areas (A) through the ``Extreme Universe'' collaboration 21H05189 and JSPS Grant-Aid for Scientific Research (20K03929). 
The work of TT is supported by JSPS KAKENHI Grant-Aid for Scientific Research (JP18K03630, JP19H01901) and for Exploratory Research (JP22K18604).

\appendix
\renewcommand{\theequation}{A.\arabic{equation}}
\setcounter{equation}{0}

\section{Ellis-Gibbons class of C-metric solutions}

Static and spherically symmetric solutions to the Einstein-phantom scalar system 
without potential 
divides into three classes \cite{Ellis1973,Martinez:2020hjm}: (i) the Fisher class~\cite{Fisher:1948yn}, (ii) the Ellis-Gibbons class~\cite{Gibbons:2003yj}  and (iii) the 
Ellis-Bronnikov class \cite{Bronnikov1973}. All of these solutions are asymptotically flat. 
Only the Fisher class solution exists even for the non-phantom case and correspond to 
the nakedly singular spacetime except for the Schwarzschild case, i.e., trivial scalar field.
The Ellis-Gibbons class solution (widely referred to as the ``exponential metric'') with a positive mass does not admit curvature singularity. 
However, it admits a parallelly propagated (p.p.) curvature singularity \cite{Hawking:1973uf} at the center, implying 
that it is not qualified as a regular wormhole \cite{Martinez:2020hjm}. Only the Ellis-Bronnikov class describes a novel wormhole geometry. 

The prescription of adding scalar potential to these solutions developed in \cite{Nozawa:2020gzz} gives rise to corresponding asymptotically 
AdS solutions with two different branches. In the non-phantom case, the AdS-Fisher solutions 
recover the solutions in \cite{Faedo:2015jqa,Anabalon:2012ta,Feng:2013tza} corresponding to hairy black holes. 
Its accelerated generalization corresponding to the C-metric has been considered in \cite{Nozawa:2022upa,Lu:2014ida,Lu:2014sza}, which 
properly accounts for the existence of two branches by virtue of C-metric flipping symmetry. 
In the body of this paper, we have discussed that this program for Ellis-Bronnikov class works out in a parallel fashion. 
This appendix gives a short report on the Ellis-Gibbons class.

Taking $\beta \to \infty$ limit, the superpotential (\ref{W}) 
reduces to 
\begin{align}
\label{}
\ma W(\phi)=-\frac{g}{2} 
e^{\frac 1{\sqrt 2}\phi } \left(
\frac 1{\sqrt 2} \phi-1 
\right) \,. 
\end{align}
The origin $\phi=0$ is an AdS vacuum corresponding to the critical point of 
the (super)potential with $\ma V(0)=-3g^2$, $\ma V''(0)=2g^2$. 
The other critical point $\phi=1/\sqrt 2$ is not the present concern.

This theory admits  the following AdS Ellis-Gibbons class of solutions \cite{Nozawa:2020gzz},
\begin{subequations}
\label{EG}
\begin{align}
\D s^2&=-e^{-2m/ r} \Delta^{\rm G}_\pm ( r)\D t^2+e^{2m/r}
\left(
\frac{\D  r^2}{\Delta_\pm^{\rm G}( r)}+ r^2 \D\Sigma_k^2(x,\varphi)
\right)\,, \\
\label{GibbonsdSstaticscalar1}
\phi=& \pm \sqrt{ \frac{-\epsilon}{2}}\frac{2m}{ r}\,.
\end{align}
\end{subequations}
Here, 
\begin{align}
\label{}
\Delta^{\rm G}_+(r)\equiv  k +g^2 r^2 e^{4m/r}\,, \qquad 
\Delta^{\rm G}_-( r)\equiv  k+g^2 (8m^2+4m  r+ r^2)\,,
\end{align}
$m$ is a parameter corresponding to the mass. 
Although curvature invariants remain finite for $m>0$ at $r=0$, this point 
corresponds to the p.p. curvature singularity.

It has been remained an open question as to why there appear two branches of solutions
for a given theory. This can be understood by the flipping symmetry of the 
C-metric. Following the strategy laid out in footnote \ref{fn:construction}, we found the following C-metric solution 
\begin{align}
\label{}
\D s^2=& \frac 1{A^2(x-y)^2}\left[ e^{-2mAx}\left(
-e^{2mA y}\Delta_y^{\rm G}(y) \D t^2+\frac{\D y^2 }{e^{2mA y}\Delta_y^{\rm G}(y)}
\right)\right.
\notag \\
&\left. 
+e^{-2mAy}\left(
e^{2mA x}\Delta_x^{\rm G}(x)\D\varphi^2 +\frac{\D x^2 }{e^{2mA x}\Delta_x^{\rm G}(x)}
\right)
\right]\,, \\
\phi=&-\sqrt 2 mA (x-y)\,,
\end{align}
with structure functions
\begin{align}
\label{}
\Delta_x^{\rm G}(x)=-c_0-c_1 x-c_2 x^2+\frac{g^2}{A^2}e^{-4mA x} \,, \qquad 
\Delta_y^{\rm G}(y)=c_0+c_1 y+c_2 y^2 \,. 
\end{align}
Following the strategy laid out in the body of text, this C-metric reduces to the minus branch 
of (\ref{EG}) in the zero acceleration limit, by the following parametrization 
\begin{align}
\label{}
c_0=-1+\frac{g^2}{A^2}\,, \qquad c_1=-\frac{4mg^2}{A} \,, \qquad c_2=k+8m^2g^2 \,.
\end{align} 
 The plus branch of the solutions is obtainable by flipping transformation (\ref{flip})
 with 
 \begin{align}
\label{}
c_0=1\,, \qquad c_1=0\,, \qquad c_2=-k \,. 
\end{align}

\end{document}